\documentclass{article}
\usepackage{amssymb}

\usepackage{graphicx}
\usepackage{amsmath}

\begin{document}

\title{From market games to real-world markets}
\author{P. Jefferies$^{1}$, M. Hart$^{1}$, P.M. Hui$^{2}$ and N.F. Johnson$^{1}$ \\
$^{1}$Physics Department, Oxford University, Oxford, OX1 3PU, U.K.\\
$^{2}$Physics Department, Chinese University of Hong Kong, Shatin, Hong Kong}
\maketitle

\begin{abstract}
This paper uses the development of multi-agent market models to present a
unified approach to the joint questions of how financial market movements
may be simulated, predicted, and hedged against.

We first present the results of agent-based market simulations in which
traders equipped with simple buy/sell strategies and limited information
compete in speculatory trading. \ We examine the effect of different market
clearing mechanisms and show that implementation of a simple Walrasian
auction leads to unstable market dynamics. \ We then show that a more
realistic out-of-equilibrium clearing process leads to dynamics that closely
resemble real financial movements, with fat-tailed price increments,
clustered volatility and high volume autocorrelation.

We then show that replacing the `synthetic' price history used by these
simulations with data taken from real financial time-series leads to the
remarkable result that the agents can collectively learn to identify moments
in the market where profit is attainable. \ Hence on real financial data,
the system as a whole can perform better than random.

We then employ the risk-control formalism of Bouchaud and Sornette in
conjunction with agent based models to show that in general risk cannot be
eliminated from trading with these models. \ We also show that, in the
presence of transaction costs, the risk of option writing is greatly
increased. \ This risk, and the costs, can however be reduced through the
use of a delta-hedging strategy with modified, time-dependent volatility
structure.
\end{abstract}

\noindent {\em Paper presented at APFA2 Conference, Liege (2000)}

\section{Introduction\label{s intro}}

Agent-based models of complex adaptive systems are attracting significant
interest across a broad range of disciplines \cite{Arthur_Overview}. \ An
important application receiving much attention within the physics community,
is the study of fluctuations in financial time-series \cite{EconPhys_Website}%
. \ Currently many different agent-based models exist in the `econophysics'
literature, each with its own set of implicit assumptions and interesting
properties \cite{Lux_Model} \cite{Cont-Bouchaud_Model} \cite
{Challet_Toy-Market} \cite{Farmer_Big-Paper}. \ In general these models
exhibit some of the statistical properties that are reminiscent of those
observed in real-world financial markets: fat tailed distributions of
returns, clustered volatility and so on. \ These models, despite their
differences draw on several of the same key ideas; feedback, frustration,
adaptability and evolution.

The Minority Game (MG) introduced by Challet and Zhang \cite
{Challet-Zhang_Original-MG} offers possibly the simplest paradigm for a
system containing these key features. \ Unlike the sophisticated model of
Lux \cite{Lux_Model} there is no external noise process simulating
information arrival. \ Nor is there any element of agents sharing local
information as in the model of Cont \& Bouchaud \cite{Cont-Bouchaud_Model}.
\ The MG simply comprises an odd number of agents \emph{N} choosing
repeatedly between the options of buying (1) and selling (0) a quantity of a
risky asset. \ The resource level of this asset is finite and therefore the
agents will compete to buy low and sell high. \ This gives the game its
`minority' nature; an excess of buyers will force the price of the asset up,
consequently the minority of agents who have placed sell orders receive a
good price at the penalty of the majority who end up buying at an
over-inflated price. \ The MG agents act with inductive reasoning, using
strategies that map the series of recent (binary)\ asset price fluctuations
to an investment decision for the next time-step. \ In an attempt to learn
from their past mistakes the agents constantly update the `score' of their
strategies and use only the most successful one to make their prediction.

The basic assumptions of this system are minimal but the resultant dynamics
show a richness and diversity that has been the focus of much recent study
\nocite{Challet-Zhang_Original-MG, Savit_Original-MG, Sherrington_TMG,
Rodgers_Modified-MG, Challet_MG-Memory, Us_Crowd-AntiCrowd}. \ However, the MG as
a realistic market model has many shortcomings:

\begin{itemize}
\item  All agents trade at each time-step

\item  All agents trade equal quantities

\item  The system resource level is fixed

\item  Agent diversity is typically limited
\end{itemize}

Many of these as well as other interesting extensions (such as agents having
the ability to learn of their own market impact \cite{Challet_Toy-Market})
have been studied separately and are discussed in \cite{EconPhys_Website}. \
This paper aims to jointly develop many of these extensions to the basic MG
in an attempt to build a minimal and yet realistic market model.

The development and study of market models from a physicist's standpoint is
motivated by the desire to learn what key interactions are responsible for
phenomena observed in the real-world system, the financial marketplace. \
However, the scope for using such market models is not simply limited to
qualitative phenomenological studies. \ The models may be extended or
manipulated to explore quantitatively the emergence of empirical scaling
laws. \ Alternatively, the approach to `critical' self-organized, or stable
states may be examined \cite{Rodgers_Model}. \ These are just a few of the
uses which could be categorized as `theoretical' study. \ What then can
these models be used for on a more `practical' or perhaps commercial level?

Recently we have been working on the possibility of using these market
models in a similar way to the way in which a meteorologist may use a model
of atmospheric dynamics; i.e. condition the models with observed data and
let them run into the future to extract probabilistic forecasts. \ These
forecasts may then be used for not only speculative gain but also for more
insightful risk management and portfolio optimization. \ Section \ref{s
Market models} of this text will expand on the idea of using the MG as a
market model, detailing the extensions needed, Section \ref{s M maker} will
then explore two different market-making mechanisms, assessing the resultant
dynamics, Section \ref{s Prediction} will detail how these models may be
used for predictive purposes and Section \ref{s Risk} will focus on risk and
portfolio optimization.

\section{The MG as a market model\label{s Market models}}

\subsection{The Basic MG}

As mentioned in the previous section, the MG formulation captures some of
the behavioral phenomena that are thought to be of importance in financial
markets; those of competition, frustration, adaptability and evolution. \ It
is also a `minimal` system of only few parameters:

\bigskip

$N=$ Number of agents

$m_{i}=$ `Memory' of agent $i$

$s_{i}=$ Number of strategies held by agent $i$

$\bigskip $

The \emph{memory} of an agent is the number of bits of the most recent past
global history that are used by a strategy in order to form a prediction. \
The agents are assigned their $s_{i}$\ strategies at the start of the game
and are not allowed to replace them at any point. \ Each agent uses the
historically most successful of her strategies to form a prediction, the
predictions of all agents are then pooled and the global history is updated
with the prediction of the minority group.

A single \emph{strategy} maps each of the $2^{m}$\ possible `histories' to a
prediction. \ Thus there are $2^{2^{m}}$\ different possible binary
strategies. \ However, many of the strategies in this space are largely
similar to one another (i.e. are separated by a small Hamming distance). It
has been shown \cite{Challet_Reduced-Space} that the principle features of
the MG are reproduced in a smaller \emph{Reduced Strategy Space} of $2^{m+1}$%
\ strategies wherein any two strategies are separated by a Hamming distance
of either 2$^{m}$\ or $2^{m-1}$\ (i.e. are \emph{anti-correlated} or \emph{%
un-correlated}). \ If the number of strategies in play i.e. $N.s$\ is
greater than $2^{m+1}$\ then the game is said to be in the \emph{crowded}'
phase, in contrast $N.s\ll 2^{m+1}$\ represents the \emph{dilute} phase.

The properties of the crowded and dilute phases of the game are quite
different and could be thought of as representing different regimes of a
market. \ In the crowded phase there will at any one time be a large number
of agents who are using the same (best)\ strategy and so will flood into the
market as large groups, producing large swings in supply and demand and a
consequently high volatility. \ If the memory of the agents is larger such
as to render $N.s\sim 2^{m+1}$\ then the groups of agents using the
same (best) strategy (\emph{crowds})\ will be smaller. \ There will also be
groups of agents who are forced to use the anti-correlated (worst) strategy,
these can be thought of as \emph{anti-crowds} as they cancel the market
action of the \emph{crowds}. \ This cancellation effect causes a reduction
in the market volatility. \ In the dilute phase it is very unlikely that any
agents will hold the same strategies and so the market behaves more randomly
and can be modelled well as a group of independent coin-tossers. \ A theory
based on these crowding effects reproduces quantitative results for the
market volatility in the basic and so called `thermal' MG across the full
range of parameters $N$,$m$,$s$. \ For more details of this the reader is
referred to \cite{Us_Crowd-AntiCrowd},\cite{Us_TMG}. \ This `Crowd,
Anticrowd Theory' may also be put to use in the formulation of an entirely
analytical set of dynamical mapping equations that reproduce the MG \cite
{Mike_Belgium}. \ These equations can be analyzed in several interesting
limiting cases to unveil the dynamics underlying microscopic behavior in
different regimes of the game. \ They may also be used in the analysis of
approaches to unstable behavior in these types of games (and possibly the
real market itself). \ Our preliminary studies have identified that there
can be at least two different `types' of build-up to a large movement (or
`crash'). \ Further work is currently underway to investigate the various
`types' of crash that can occur and their precursors.

\subsection{The Grand-Canonical MG}

In the basic MG agents must either buy or sell at every time-step. \ In a
real market however, traders are likely to wait on the sidelines until they
are reasonably confident that they are able to make a profit with their next
trade. \ They will observe the market passively, mentally updating their
various strategies, until their confidence overcomes some threshold value -
then they will jump in and make a trade. \ We now demonstrate an extension
to the basic MG which attempts to incorporate this general behavioral
pattern.

The primitive binary agents of the basic MG keep a tally of the \emph{virtual%
} score $r_{S,i}$\ of each of their $s_{i}$ strategies: $+1$ for a correct
prediction and $-1$ for an incorrect prediction, and \emph{virtual} in the
sense that the strategy is scored whether it is played or not. \ They may
also keep a tally of their own personal prediction success score $r_{i}$. \
It is reasonable that each agent $i$ has a finite time horizon $T_{i}$ over
which these success scores are monitored; this is equivalent to a `sunken
losses' approach. \ We now make the simplest possible generalization which
is to introduce a threshold value $r_{\min }$\ in either $r$ or $r_{S}$
below which an agent would choose to not trade. \ In this case, the agent
continues to update her strategy scores $r_{S}$ but now adds a $0$ to her
personal score tally $r$. \ With this extension, the number of agents
actively trading at each time-step $N_{active}$ will vary throughout the
game. \ This feature is reminiscent of the Grand-Canonical-Ensemble in
statistical mechanics.

If an agent's threshold to play lies at the lower end of the range $-T\leq
r_{\min }\leq T$ then we would expect the agent to play a large proportion
of the time as her best strategy will have invariably scored higher than
this threshold. \ Conversely, for high $r_{\min }$, the agent will scarcely
play at all. \ We would thus expect to see a transition occur between these
two regimes at intermediate values of the threshold. \ Figure 1 shows the
time-averaged number of active agents $\left\langle N_{active}\right\rangle $
and the standard deviation of this quantity as a function of $r_{\min }$ for
a uniform population of $N=101$ $m=2$, $s=2$ agents who record scores over $%
T=50$ time-steps. \ Here $r_{\min }$, the threshold to play, is based on the
agent's strategy score $r_{S,i}$ such that an agent only plays if $\max %
\left[ \left\{ r_{S,i}\right\} \right] >r_{\min }$. \ A similar transition
effect is also seen if the threshold is based on prediction success score $%
r_{i}$.

The behavior of $\left\langle N_{active}\right\rangle $\ can be reproduced
to a coarse approximation by assuming that the strategy scores $r_{S,i}$
undergo independent binomial random walks: 
\begin{equation*}
r_{S}\thicksim 2Bin\left[ T,\frac{1}{2}\right] -T
\end{equation*}
This gives: 
\begin{eqnarray*}
\left\langle N_{active}\right\rangle &\thickapprox &N\left( 1-P\left[
r_{S}<r_{\min }\right] ^{s}\right) \\
\sigma ^{2}\left[ N_{active}\right] &\thickapprox &N\left( 1-P\left[
r_{S}<r_{\min }\right] ^{s}\right) P\left[ r_{S}<r_{\min }\right] ^{s}
\end{eqnarray*}
This approximation captures the essence of the transition mentioned in the
paragraph above. \ However, the behavior of $r_{S}$ is in reality far from
that of a random walk. \ In the crowded regime $r_{S}$ is strongly
mean-reverting and in the dilute regime of the game it has a strong drift
component, also the increments in individual strategy scores can be highly
correlated. \ The approximation becomes better for $T\gg 2^{m}$, where many
of these effects become averaged out.

With intermediate values for $r_{\min }$ this modified MG produces very
interesting dynamics \cite{Neil_Dublin}, for instance there can be moments
of extreme illiquidity followed by a rush to the market causing huge swings
in supply and demand. \ There are also noticeable `ranging' and `break-out'
periods and other patterns familiar to market traders \cite{Us_B1-Work}.

We now extend this model to allow $r_{\min }$\ to be dynamic. \ Here each
agent decides on her own threshold in a manner dependent on her current
internal state variables. \ This allows an enhanced element of evolution
within the model and more closely resembles behavioral models of markets
wherein levels of confidence are time-dependent. \ We choose to make $%
r_{\min }$\ a function of the agent's personal success rate $r_{i}$. \
Asserting that agents are rational and risk-averse implies that $r_{\min }>0$
and that $\frac{dr_{\min ,i}}{dr_{i}}\leq 0$ i.e. never play a strategy that
has lost more times than won and take fewer risks if losing. \ Following
basic utility theory we therefore arrive at: $r_{\min ,i}=\max \left[
0,-\left( r_{i}-\lambda .\sigma \left[ r_{i}\right] \right) \right] $ (where 
$\sigma \left[ r_{i}\right] $\ is the player's standard deviation of success
and $\lambda $\ is their coefficient of risk-aversion). \ As agents' success
rates vary in time, then so will their threshold values and we see an
overall evolution towards a diverse population as shown in Figure 2.

This version of the `Grand-Canonical' MG forms the basic framework for our
development of a market model. \ The following subsection will outline the
further necessary extensions to the model that, when combined, form our
`realistic' market model.

\subsection{Agent diversity \& wealth}

It is a simple extension of the model developed above, to include
agent heterogeneity in terms of wealth, investment size and investment
strategy. \ As it stands, each trade made by an agent is the exchange of one
quanta of a riskless asset for one quanta of a risky one, irrespective of
the agent's wealth or the price of the asset. \ Also, agents always trade as
`value' investors, seeking to buy low and sell high at each time-step. \ We
now generalize this framework to introduce a more realistic heterogeneity
between investors.

We first allot each agent $i$\ a quantity of each asset, riskless $B_{i}%
\left[ 0\right] $ and risky $S_{i}\left[ 0\right] $. \ When a trade is made,
it is made at the market price of $p\left[ t\right] \pm \delta \left[ t%
\right] $ where $\delta \left[ t\right] $\ corresponds to a spread raised by
the marketmaker (the market-making mechanism is the subject of the next
section). \ We now re-assert the assumption that investors are risk-averse
and will thereforee trade amounts proportional to their absolute wealth. \
We also assume that the amount they trade will be in proportion to their
confidence in the strategy they intend to use. \ It is helpful at this stage
to define a measure of this confidence $c_{i}$. 
\begin{equation*}
c_{i}\left[ t\right] =\frac{\max \left[ r_{S,i}\left[ t\right] \right]
-r_{\min }\left[ t\right] }{T_{i}}
\end{equation*}
thus $-2<c_{i}<1$ but the agent only plays if $c_{i}>0$. \ Buy operations
are then represented by\bigskip : 
\begin{eqnarray*}
B_{i}\left[ t+1\right] &=&B_{i}\left[ t\right] \left( 1-c_{i}\left[ t\right] 
\frac{p\left[ t+1\right] +\delta \left[ t+1\right] }{p\left[ t\right]
+\delta \left[ t\right] }\right) \\
S_{i}\left[ t+1\right] &=&S_{i}\left[ t\right] +\frac{c_{i}\left[ t\right]
B_{i}\left[ t\right] }{p\left[ t\right] +\delta \left[ t\right] }
\end{eqnarray*}
and sell operations by: 
\begin{eqnarray*}
B_{i}\left[ t+1\right] &=&B_{i}\left[ t\right] +c_{i}\left[ t\right] S_{i}%
\left[ t\right] \left( p\left[ t+1\right] -\delta \left[ t+1\right] \right)
\\
S_{i}\left[ t+1\right] &=&S_{i}\left[ t\right] \left( 1-c_{i}\left[ t\right]
\right)
\end{eqnarray*}
Wealthy agents make large transactions and thus will have a high market
impact (in a system where price movement size is an increasing function of
order size, c.f. Equation \ref{Eq Price set Farmer}) whereas poor agents
effectively form a background `noise' of small trades. \ Of course poor
agents may grow rich or vice-versa. \ When agents have lost all their
assets, they can no longer trade; this represents the bankruptcy of that
agent. This situation happens extremely rarely in these models and so we
have not sought to implement a system for the re-generation of new agents. \
Figure 3\ shows the average distribution of agents' wealth as measured by $B%
\left[ t\right] +S\left[ t\right] .p\left[ t\right] $ (i.e. the
probabilities are averaged over time).

As well as the diversity in agents' trade size, there can also be a
diversity in investment strategy. \ Within the framework presented here,
investment strategies can fall into the two broad classes; \emph{value} and 
\emph{trend}. \ A \emph{value} investor aims at each time-step to make a
profit from buying low and selling high, a \emph{trend} investor on the
other hand aims to buy an upward moving asset and sell a downward mover. \ A
population purely of value investors will have a minority-game character, a
population of trend investors will create a majority-game of self-fulfilling
prophecies. \ In general, the population of traders will be a combination of
these types and thus the character of the market (minority or majority) is
unclear. \ We are currently testing how the proportion of each investor type
alters the global dynamics and stability of the market.

\section{The Market-Making Mechanism\label{s M maker}}

\subsection{Walrasian auction}

The simplest type of market-making process is that of a walrasian auction.
This is a popular model in the economics community (and actually the system
used in the London Metals Exchange). \ In a walrasian auction investors take
part in a price setting process by submitting orders to buy or sell the
risky asset based on a theoretical price. \ The level of this theoretical
price is changed until the supply and demand for the asset exactly match and
the market can be cleared, then the process repeats.

We can use our market model to simulate a simplified version of this process
in the following way. \ First of all we assume that the supply and demand
are in equilibrium at each time-step. \ The resulting equilibrium price for
the risky asset then must be equal to the current demand-value of stocks
sought divided by the number of risky assets offered. \ This gives: 
\begin{equation*}
p[t+1]=\frac{\sum\limits_{i,\text{Buyers}}c_{i}\left[ t\right] B_{i}\left[ t%
\right] }{\sum\limits_{i,\text{Sellers}}c_{i}\left[ t\right] S_{i}\left[ t%
\right] }
\end{equation*}
It is clear then that this process is unstable: if there are no buyers the
price falls to zero and if there are no sellers it will rise to infinity! \
Even though these situations happen rarely in a run of the market model, the
resulting dynamics still show an inherent instability and the fluctuations
are excessive as well as exhibiting a strong anti-persistence. \ This
situation arises because we are asserting that the buy and sell pressures
are in equilibrium at each time-step. \ Of course this is far from the
reality and we must extend the market-making mechanism to accommodate the
real out-of-equilibrium process.

\subsection{Non-equilibrium market}

If the supply of risky assets does \textbf{not} exactly match the demand at
each time-step then the market will either not clear, or the market-maker
will take a position in the asset himself in order to fill the orders. \ In
reality it is most likely that a combination of these scenarios occurs; the
market maker will want to fill as many orders as possible and take the
spread but he will not allow himself to incur a large position. \ There are
many ways in which this type of behavior could be modelled, we limit
ourselves here to looking at one particular system.

Let us start by implementing the price setting rule of Bouchaud - Cont -
Farmer \cite{Cont-Bouchaud_Model},\cite{Farmer_Big-Paper} : 
\begin{equation}
p\left[ t+1\right] =p\left[ t\right] e^{\frac{Buys\left[ t\right] -Sells%
\left[ t\right] }{Liquidity}}  \label{Eq Price set Farmer}
\end{equation}
where: $
\begin{array}{c}
Buys \\ 
Sells
\end{array}
\left[ t\right] =\sum\limits_{i 
\begin{array}{c}
\text{Buyers} \\ 
\text{Sellers}
\end{array}
}S_{i}\left[ t+1\right] -S_{i}\left[ t\right] $, and $liquidity$ is a
constant set by the market-maker. \ This rule prevents the market-maker
being arbitraged but leaves his inventory ($B_{M}\left[ t\right] $ and $S_{M}%
\left[ t\right] $) as unbounded. \ Over many runs of such a market
simulation we would expect the market-maker's mean inventory to be zero.
What we really require on the other hand is that his mean inventory in a
particular run be zero. \ We therefore propose the following extension to
Equation \ref{Eq Price set Farmer}: 
\begin{equation}
p\left[ t+1\right] =p\left[ t\right] e^{\frac{Buys\left[ t\right] -Sells%
\left[ t\right] -S_{M}\left[ t\right] }{Liquidity}}  \label{Eq Price set Me}
\end{equation}
\qquad
This implies that if the market-maker is accruing a net long position in the
risky asset, he'll start lowering the price in order to attract buyers into
the market and vice-versa. \ This mechanism works remarkably well and we
find that $S_{M}\left[ t\right] $ under this new rule is strongly
mean-reverting as shown in Figure 4.

With Equation \ref{Eq Price set Me} however the market maker can be
arbitraged by the agents; the strategy buy, wait, sell or vice-versa will
make money as long as enough agents do it at the same time. \ The agents in
these systems learn to exploit this very quickly (an interesting result in
itself) and the result is a negative drift to the market-maker's money $B_{M}%
\left[ t\right] $. \ There are several mechanisms that the market-maker may
exploit to overcome this; he can raise a spread or he can reduce the
liquidity. \ We employ the first of these mechanisms, updating the spread
proportionally to $-\frac{\left\langle B_{M}\left[ t\right] \right\rangle }{%
\left\langle v\left[ t\right] \right\rangle }$ where $v\left[ t\right] $ is
the volume of transactions defined as $v\left[ t\right] =\sum%
\limits_{i=1}^{N}S_{i}\left[ t+1\right] -S_{i}\left[ t\right] $ . \ Here
$\left\langle B_{M}\left[ t\right] \right\rangle $ and $\left\langle v%
\left[ t\right] \right\rangle $ are taken over a time-length $T_{M}$ which
is kept large compared with $\left\{ T_{i}\right\} $ such as to average over
local extreme behavior such as momentary illiquidity. \ This mechanism for
raising a spread may not be highly efficient but it does maintain the
market-maker's mean wealth close to the desired zero point by raising the
spread if he starts losing money. \ The $1/v\left[ t\right] $ dependence
stabilizes this process somewhat by sharing the job of paying for the
market-maker's deficit over the current number of market participants.

We now have a complete and arguably `realistic' model and may begin to
investigate its properties. \ We are at present looking at how different
statistical properties of the model-market are dependent on its parameters.
\ We seem to find however that some statistical features are in general
present over very large parameter ranges. \ These are the types of feature
that are associated with `real' markets: High excess kurtosis of returns
with weak decay over time, volatility clustering, high volume
autocorrelation etc. as shown in Figure 5.

\section{Prediction from market-models\label{s Prediction}}

The market-models introduced in Section \ref{s Market models} consist of a
population of adaptive agents who attempt to predict the future movement of
an asset price. \ Recently, we have been investigating the accuracy of these
predictions when the synthetic self-generated global history of asset
movements is replaced with a real financial time series.

The first step in this process is to generate binary information from the
given financial time-series. \ This can be done in many ways in order to
investigate the predictability of different aspects of the movement. \ We
choose here to examine the sign of movements and hence our information
history $h\left[ t\right] $ becomes: 
\begin{equation*}
h\left[ t\right] =H\left[ p_{\text{real}}\left[ t\right] -p_{\text{real}}%
\left[ t-1\right] \right] 
\end{equation*}
where $H\left[ x\right] $ is the Heaviside function. \ If $p_{\text{real}}%
\left[ t\right] =p_{\text{real}}\left[ t-1\right] $ then we assign $h\left[ t%
\right] $ a $0$\ or $1$ randomly. \ Before we begin to look at how the
agent-models perform with this new information set, let us first examine
some of its properties. \ The agents examine chunks of the information set $h
$ of length $m$ bits in order to make a prediction. \ If we look at the
occurrence rate of $m+1$ length bit-strings we can therefore infer the
success rate of strategies. \ For example Figure 6\ shows the occurrence
probability of 4-bit strings; i.e. 3 memory bits ($m=3$) and one prediction
bit. The bit-strings are enumerated by their decimal value e.g. $%
0011\rightarrow 3$. \ We can infer that the strategy $\left\{
10101010\right\} $ (which is the $m=1$, $\left\{ 10\right\} $ i.e.
anti-persistent strategy)\ will have the highest success rate as $000$ is
more often followed by a $1$, $001$ by a $0$ etc.

As we decrease the sampling rate on our data-set so as to look at the signs
of price increments over longer periods, we find that the most successful
strategy becomes less well defined and tends to swap regularly. It is no
longer the case that a simple anti-persistent strategy is the best. \ Also
as we increase the memory $m$ and look at longer bit-strings, we find that
the `information content' of the bit-string occurrence histograms gets
`washed away' in the mixing of low $m$ probabilities. This implies that the
most dominant physical process is a low $m$ process. \ Figure 7\ shows these
two effects by examining the excess standard deviation of the bit-string
distributions i.e. $\sigma _{Bitstring}^{\text{real}}/\sigma _{Bitstring}^{%
\text{random}}$ where $\sigma _{Bitstring}^{\text{random}}=\frac{\sqrt{%
L\left( 2^{m+1}-1\right) }}{2^{m+1}}$, where $L$ is the length of the
data-set.

In the agent simulations, $h\left[ t\right] $ plays the same role as before,
with strategies and agents being scored for prediction success in the same
fashion as detailed in Section \ref{s Market models}. \ Of course now the
feedback has been removed from the model, it bears more resemblance to a
system of genetic algorithms. \ The key important difference though is the
fact that this system of independent agents has a large built in
frustration: the agents aren't allowed to replace poorly performing
strategies. \ Although this at first appears to be a handicap, it can in
fact be a strength. \ In systems where there is not necessarily a `correct'
strategy to employ, there is an advantage in having many currently
non-optimal strategies in play as this gives greater adaptability. \ We have
compared the prediction success of these types of model with those employing
simple Bayesian update of the probability of a given outcome for a given
history and found the former to be much more powerful. \ Figure 8 shows the
time-series of the \$/Yen FX-rate between 1990-99, below this is a plot of the
cumulative non-compounded profit attained from using the agent model's
predictions to trade hourly. \ The trading strategy employed is simply to put the
original investment amount on either the \$ or the Yen side of the market and
take it off again at the end of the hour, banking the profit in a zero interest
account. \ This is clearly an unrealistic strategy as transaction costs would be
penalizing, however it is used in order to demonstrate simply that the
agent-model performs better than random (around 54\% prediction success rate)
\footnote{We have run these models with randomly generated information
histories $h \left[ t\right] $ and were able to reject the null hypothesis that
the mean prediction success rate with real data was random.}. \ The two profit
lines on Figure 8\ represent two different uses of the independent predictions of
the agents. \ The lower line corresponds to the case where the investment is
split equally between all agents, the upper line is for the case where the
agents' predictions are pooled together with a non-linear function. \ This
demonstrates that the population as a compound entity can perform much
better than the sum of its individual parts. This kind of phenomena has been
termed `collective intelligence' in the past.

Arguably the most interesting phenomena of models such as the MG arise from
the strong feedback mechanism. \ In replacing the self-generated $h\left[ t%
\right] $ with an external process we disable that feedback. \ The system is
still however able to function as a weak predictor. \ It appears that the
prediction success rate can be raised by invoking again a feedback within
the system. It is probable that this feedback forces a more efficient
learning process to take place. \ These effects are the subject of our
current, ongoing studies.

We have hence demonstrated the success of the agent-based models in direct
prediction of the sign of the next price increment. \ However, we can also
implement the models in a different way by `training' them on historical
data of a particular asset movement and then using the artificial
market-making process to run the models forward into the future. \ If this
is done with an ensemble of such models, each having a different initial
allocation of strategies, we can form a distribution of likely future asset
price levels. \ Typically the resulting distributions are fat tailed and can
have considerable skewness quite in contrary to more standard economic
models. \ This information can not only be of use in speculation but also in
risk control and portfolio management.

\section{Risk management\label{s Risk}}

\subsection{Implied future risk from agent-models}

The control of risk in financial investment should be of equal importance to
the realization of profit. \ Most current theories of risk control rely on
the implicit assumption that future behavior of the market will be like its
past behavior. \ This assumption is continually being brought into question
when banks and investors seem to be `caught out' by events that past
distributions seemed to imply were impossible. \ There thus may be room here
for risk-control models that rely more on possible emergent future behavior
than on historic data.

Using agent-based models in the way mentioned at the close of Section 4 gives us
distributions for likely future price levels based on what microscopically might
happen. \ This may be just the type of forward-casting model that could be of use
here. \ We must first however develop a framework within which we can use the
type of information that these models give us. \ Much of risk-control concerns
itself with the use of derivative instruments, we therefore follow this direction
but take pause to note that a similar methodology can be used for analyzing any
portfolio of assets.

Several years ago Bouchaud and Sornette developed a framework for examining
and controlling the risk inherent in writing derivative contracts \cite
{Bouchaud_Deriv-Formalism}. \ This formalism explicitly deals with future
asset movements in a probabilistic, path-dependent fashion i.e. does not
rely on any random-walk model etc. \ This makes the formalism ideal for
combining with the forward-casting agent-models.

The formalism examines the variation in future wealth $\Delta W_{T}$\ from
holding a certain portfolio, for example short one euro-call contract of
price $C_{0}$\ maturity $T$ and strike $X$ and long $\phi _{t}\left[ S_{t}%
\right] $ hedging assets in the underlying which is at price $S_{t}$ at time 
$t$: 
\begin{equation}
\Delta W_{T}=C_{0}-\max \left[ S_{T}-X,0\right] +\sum_{t=0}^{T}\phi _{t}%
\left[ S_{t}\right] \left( S_{t+1}-S_{t}\right)\ .  \label{eq Bouch VariW}
\end{equation}
The variance\ of this wealth process (which is used as a measure of risk) is
then found analytically for a general underlying movement. \ For our models,
this can be done in a Monte-Carlo fashion using each member of the model
ensemble to generate a $\Delta W_{T}$. \ Doing this we could also look at
other measures of risk such as VAR etc. \ This process generates a more
insightful measure of risk based on likely future microscopic behavior.

The control of this risk is the next issue. \ Bouchaud and Sornette's
variance of the wealth process can be minimized with respect to the hedging
strategy $\phi _{t}\left[ S_{t}\right] $. \ The full details are given in 
\cite{Bouchaud_Risks-Book}; the result is a risk-minimizing `optimal
strategy' given by: 
\begin{equation}
\phi _{t}\left[ S_{t}\right] =\int_{X}^{\infty }\frac{\left(
S_{T}-S_{t}\right) \left\langle \delta S_{S_{t},t\rightarrow
S_{T},T}\right\rangle }{\left\langle \delta S_{t}^{2}\right\rangle }P\left[
S_{T}|S_{t}\right] dS_{T}  \label{eq Bouch opStrat}
\end{equation}
Using the forward-casting agent-models we obtain $P\left[ S_{T}|S_{t}\right] 
$\ (the probability of the underlying moving from value $S_{t}$ to $S_{T}$)
by counting the number of members of the (large) model ensemble that cast
paths passing near both these two values (price space $S$ is discretized for
this purpose). \ Similarly $\left\langle \delta S_{S_{t},t\rightarrow
S_{T},T}\right\rangle $ is found as the mean increment at time $t$ of paths
passing near $S_{t}$ and $S_{T}$, $\left\langle \delta
S_{t}^{2}\right\rangle $ is simply the mean squared increment at time $t$\
of all paths. \ The resulting reduction in risk when using this `optimal
strategy' with historical distributions is well documented \cite
{Bouchaud_Risks-Book}; similar effects are obtained when using the agent-models'
future-cast distributions. \ The important difference to note is that the
risk being minimized is now the microscopically derived future risk rather than
a measure assuming the continuity of past behavior.

\subsection{Transaction costs}

We now digress slightly and examine the effect of transaction costs on the
risk control process discussed in the previous paragraphs. \ Bouchaud and
Sornette's formalism is easily couched in discrete time, accounting for the
fact that continuous trading is un-physical due to transaction cost and
brokerage inefficiencies. \ However, transaction costs themselves have not
explicitly been accounted for in the wealth process, therefore their effect
on risk-control cannot be gauged. \ We address this point here by adding a
term to equation \ref{eq Bouch VariW} in order to include a general
transaction cost structure. 
\begin{equation*}
\Delta W_{T}\rightarrow \Delta W_{T}+\sum_{t=0}^{T}k_{1}+\left(
k_{2}+k_{3}S_{t}\right) \left| \phi _{t}\left[ S_{t}\right] -\phi _{t-1}%
\left[ S_{t-1}\right] \right|
\end{equation*}

We now again proceed to find the variance of this wealth process as a gauge
of risk. \ We find that the approximation of $\left| \phi _{t}\left[ S_{t}%
\right] -\phi _{t-1}\left[ S_{t-1}\right] \right| \approx \frac{\partial
\phi _{t}}{\partial S_{t}}\left| \delta S_{t}\right| $ holds reasonably well
as the time dependence of $\phi _{t}\left[ S_{t}\right] $ is weak. \ This
allows us to formulate an analytical correction term to Bouchaud and
Sornette's expression for risk (full details will be presented elsewhere): 
\begin{align}
R& \rightarrow R+\sum_{t=1}^{T}\left( 
\begin{array}{c}
\int_{-\infty }^{\infty }\left\langle \delta S_{t}^{2}\right\rangle \left(
k_{2}+k_{3}S_{t}\right) ^{2} \\ 
\times \left( \frac{\partial \phi _{t}}{\partial S_{t}}\right) ^{2}P\left[
S_{t}|S_{0}\right] dS_{t} \\ 
-\left( 
\begin{array}{c}
\int_{-\infty }^{\infty }\left\langle \left| \delta S_{t}\right|
\right\rangle \left( k_{2}+k_{3}S_{t}\right) \\ 
\times \frac{\partial \phi _{t}}{\partial S_{t}}P\left[ S_{t}|S_{0}\right]
dS_{t}
\end{array}
\right) ^{2}
\end{array}
\right)  \label{Eq TC Risk} \\
& +\sum_{ti\neq tj}\iint\limits_{-\infty }^{\infty }\left( 
\begin{array}{c}
\left\langle \left| \delta S_{ti}\right| \right\rangle \left\langle \left|
\delta S_{tj}\right| \right\rangle \\ 
\times \left( k_{2}+k_{3}S_{ti}\right) \left( k_{2}+k_{3}S_{tj}\right) \\ 
\times \frac{\partial \phi _{ti}}{\partial S_{ti}}\frac{\partial \phi _{tj}}{%
\partial S_{tj}}P\left[ S_{ti}|S_{0}\right] \\ 
\times \left( P\left[ S_{tj}|S_{ti}\right] -P\left[ S_{tj}|S_{0}\right]
\right)
\end{array}
\right) dS_{ti}dS_{tj}  \notag
\end{align}
The first line of equation \ref{Eq TC Risk} represents the sum of
independent transaction cost variances whereas the second line represents
the covariance between transaction costs. \ The covariance terms become very
large as we execute more transactions. This non-local behavior leads to a
divergence of the risk as we go toward continuous time as shown in Figure 9.
\ Clearly if we are to minimize risk now the answer is not to simply
re-hedge more often.

The minimization of risk with respect to a choice of hedging strategy $\phi
_{t}\left[ S_{t}\right] $ is now highly complex and in general
path-dependent as might be expected from equation \ref{Eq TC Risk}. \
However, we may use pertubation theory to obtain approximate solutions. \ We
find that the risk and transaction costs are reduced greatly using a
volatility correction to equation \ref{eq Bouch opStrat} of the form: 
\begin{equation*}
\left\langle \delta S_{t}^{2}\right\rangle \rightarrow \gamma \left[ t\right]
\left\langle \delta S_{t}^{2}\right\rangle 
\end{equation*}
The form of $\gamma \left[ t\right] $ as a function of time is amusingly
that of a smile, much like the volatility correction in strike price to the
Black-Scholes delta that is implied by equation \ref{eq Bouch opStrat}
itself. \ The origins of these two `volatility smiles' are of course very
different. \ Using this correction, for portfolios where transaction costs
are likely to be high, we see a dramatic reduction in the risk and also in
the absolute transaction costs. \ Figure 10\ demonstrates this for a
particular option.

\section{Conclusion}

We have presented here a development from the basic minority game, to a full
market model. \ We have attempted to capture the behavioral aspects of
market-making and agent-participation in a thorough and yet simplistic
fashion. \ From this model we have then shown behavior reminiscent of `real'
financial asset movements with fat-tailed distributions of returns,
clustered volatility and high volume autocorrelation.

We then moved on to show how these types of agent-based models perform in a
predictive capacity when we replace the self-generated synthetic asset-price
history with a real financial asset movement. \ We showed that as
independent entities, the agents were able to function in a manner similar
to an inefficient genetic algorithm and thus exploit the residual
information present in the asset movement's sign. \ We then went on to show
that when combined as a population, the agents were able to perform as a
much stronger predictor, suggesting an element of collective-intelligence. \
We then outlined another manner in which ensembles of these models can be
used to forecast future asset-price levels in a probabilistic manner.

Lastly, we showed how output from the agent-models could be used in a
portfolio management setting in order to measure and control risk. \ We went
on to demonstrate that the addition of transaction costs to Bouchaud and
Sornette's formalism for risk management led to a greatly increased risk for
high-frequency trading. \ We then presented a volatility correction to the
`optimal strategy' that could be used to reduce this excess risk and also
reduce transaction costs.

Our aim is to develop a general understanding and framework for
investigating and exploiting financial markets based on microscopic models
of agent interactions. \ It is hoped that the work presented here represents
positive and significant steps toward this goal.

\bigskip

We thank A. Short for many useful discussions.

\pagebreak 

\begin{description}
\item  Figure 1:\qquad Mean and standard deviation in the number of active
agents $N_{active}$ (game parameters $N=101$, $m=2$, $s=2$, $T=50$)

\item  Figure 2:\qquad Distribution of threshold values $r_{\min }$ after
6000 time-steps (game parameters $N=151$, $m=3$, $s=2$, $T=50$, $\lambda
=0.07$)

\item  Figure 3:\qquad Time averaged PDF of agent's wealth as measured by $B%
\left[ t\right] +S\left[ t\right] p\left[ t\right] $ (Game parameters $N=151$%
, $m=3$, $s=2$, $T=50$, $\lambda =0.07$). Original allocation of wealth; $B%
\left[ 0\right] =1000\$$, $S\left[ 0\right] =100$, $p\left[ 0\right] =10\$$.

\item  Figure 4:\qquad Market-maker's stock $S_{M}\left[ t\right] $ over
6000 turns (Total stock in market 15100)

\item  Figure 5:\qquad Price \& Volume Statistics for a single run of the
market simulation (parameters $N_{value}=101$, $N_{trend}=50$, $m=3$, $s=2$, 
$T=20$, $\lambda =0.07$)

\item  Figure 6:\qquad Occurence probability of 4-bit strings in the
price-sign history $h\left[ t\right] $ generated from 10 years of hourly
\$/Yen FX-rate data

\item  Figure 7:\qquad $\sigma _{Bitstring}^{\text{real}}/\sigma
_{Bitstring}^{\text{random}}$ as a function of price increment length for $%
m=2,5,8$ (dataset \$/Yen FX-rate between 1990-99)

\item  Figure 8:\qquad \$/Yen FX-rate 1990-99 (top) and cumulative
non-compounded profit from using agent predictions both independently and
collectively (bottom)

\item  Figure 9:\qquad Standard deviation of the wealth process (risk) as a
function of trading time (length of time between trades). \ 30-day
at-the-money european option vol=7.37p/day.

\item  Figure 10:\qquad Simulated distribution of wealth for portfolio short
one 30-day euro-call, at the money, vol=7.37p/day and long $\phi _{t}\left[
S_{t}\right] $ of the underlying with transaction costs at $k_{3}=5\%$. \ $%
\phi _{t}\left[ S_{t}\right] $ according to Black-Scholes Delta (top) and
with modified volatility as described in the text (bottom).
\end{description}

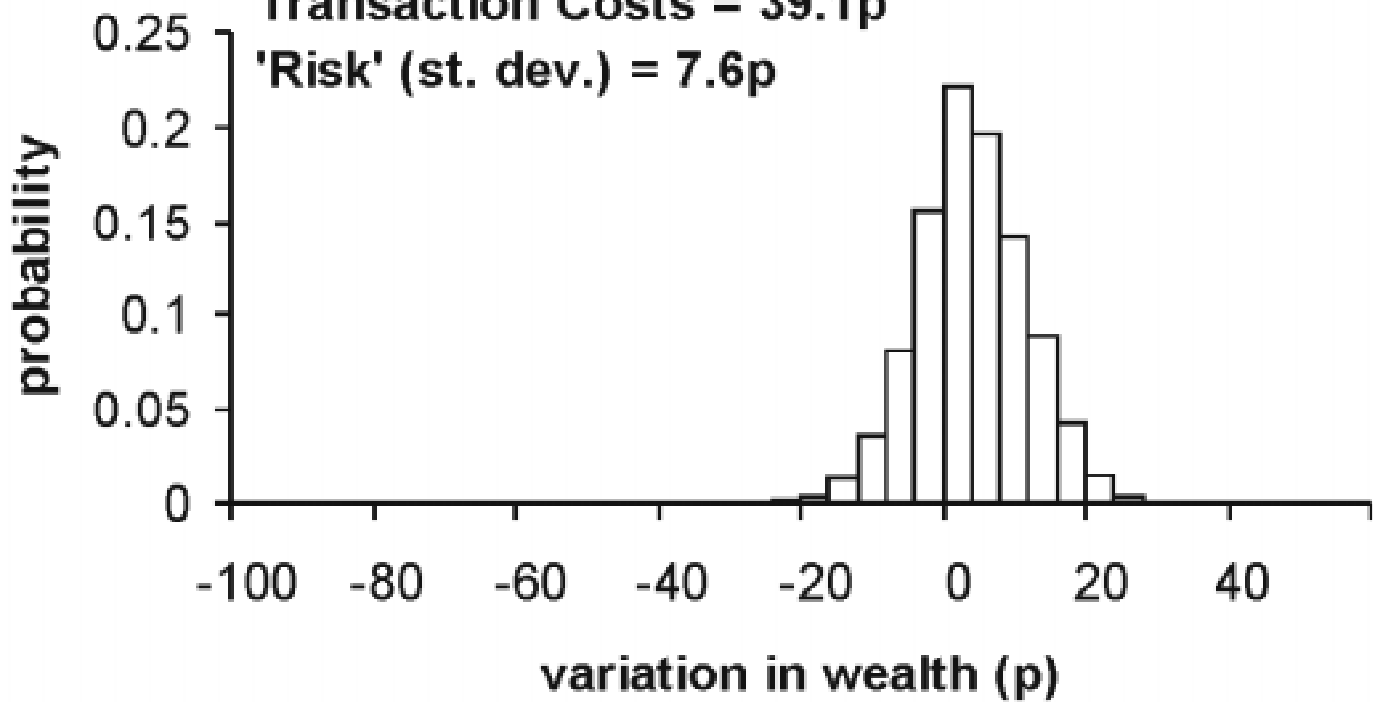
\end{document}